# Mechanistic Insights into Non-Adiabatic Interband Transitions on a Semiconductor Surface Induced by Hydrogen Atom Collisions

Lingjun Zhu,[1,2] Qijing Zheng,[1,3] Yingqi Wang,[4] Kerstin Krüger,[5] Alec M. Wodtke,[5,6,7] Oliver Bünermann,[5,6,7] Jin Zhao,[1,3*] Hua Guo,[4*] and Bin Jiang[1,2*]

[1]Key Laboratory of Precision and Intelligent Chemistry, University of Science and Technology of China, Hefei, Anhui 230026, China.

[2]Department of Chemical Physics, University of Science and Technology of China, Hefei, Anhui 230026, China.

[3]Department of Physics, University of Science and Technology of China, Hefei, Anhui 230026, China.

[4]Department of Chemistry and Chemical Biology, Center for Computational Chemistry, University of New Mexico, Albuquerque, New Mexico 87131, USA.

[5]Institute of Physical Chemistry, Georg-August University, Göttingen 37077, Germany.

[6]Department of Dynamics at Surfaces, Max-Planck-Institute for Multidisciplinary Sciences, Göttingen 37077, Germany.

[7]International Center of Advanced Studies of Energy Conversion, Georg-August University, Göttingen, Germany.

*: corresponding authors: zhaojin@ustc.edu.cn, hguo@unm.edu, bjiang@ustc.ac.cn

**Abstract**

To understand the recently observed mysterious non-adiabatic energy transfer for hyperthermal H atom scattering from a semiconductor surface, Ge(111)*c*(2×8), we present a mixed quantum-classical non-adiabatic molecular dynamics model based on time-dependent evolution of Kohn-Sham orbitals and a classical path approximation. Our results suggest that facile non-adiabatic transitions occur selectively at the rest atom site, featuring excitation of valance band electrons to the conduction band, but not at the adatom site. This drastic site specificity can be attributed to the changes of the local band structure upon energetic H collisions at different surface sites, leading to transient near-degeneracies and significant couplings between occupied and unoccupied orbitals at the rest atom, but not at the adatom. These insights shed valuable light on the collisional induced non-adiabatic dynamics at semiconductor surfaces.

**I. Introduction**

The Born-Oppenheimer approximation (BOA),[1] which assumes much faster electronic motion than nuclear motion, represents a key construct for understanding spectroscopy and collision dynamics. This approximation is however known to break down when two or more electronic states are energetically close to each other. For gas phase systems, where electronic degeneracies are infrequent, the breakdown of BOA is reasonably well understood and various theoretical methods for treating non-adiabatic dynamics have been developed.[2-7] In solids and at surfaces, however, the mechanisms for the breakdown of BOA remain largely unexplored until recently.

Recent experiments have started to examine non-adiabatic processes on metal surfaces, in which a large number of electronic states exist with infinitesimal energy differences. As a result, the breakdown of BOA is found to be more prevalent.[8-9] Indeed, a range of electronically non-adiabatic phenomena, such as efficient vibrational quenching,[10-11] electron emission,[12] chemicurrents,[13] and collisional energy dissipation,[14] have been reported when atoms/molecules interact with metal surfaces. These behavior can often be interpreted in terms of instantaneous formation of electron-hole pairs (EHPs) due to the coupling of metallic electrons with nuclear motion through formation of a transient anion, or arising from an electronic frictional (EF) force on nuclear motion.[15] The easy breakdown of BOA on metal surfaces is in sharp contrast with their insulator counterparts, in which the large band gap prevents electronic excitation between the valence band (VB) and conduction band (CB).[14, 16]

Very recently, non-adiabatic effects have been reported on semiconductor surfaces.

As an intermediate case between insulators and conductors, semiconductors have a significant energy gap between VB and CB that cannot be easily overcome by thermal excitation. By impinging hyperthermal H/D on a reconstructed Ge(111)*c*(2×8) surface, Krüger et al. measured the energy loss of the scattered projectiles at different incidence energies.[17] For an incidence energy below the band gap, there is only one peak in the energy loss distribution. This distribution, along with the corresponding angular distribution of the scattered H, was accurately reproduced by adiabatic molecular dynamics,[17] signaling its mechanical nature. When the incidence energy is above the bandgap, however, two peaks were found. Again, one peak can be successfully modelled with adiabatic molecular dynamics; however, the second peak exhibiting a large energy loss cannot be accounted for by any existing model. Interestingly, the threshold coincides with the surface band gap (0.49±0.03 eV[18]), prompting the conclusion that the large energy loss peak is due to non-adiabatic excitation of substrate electrons induced by the energetic collision of the fast atomic hydrogen.[17] A recent work with deuterium atoms scattering from the same surface revealed the same energy loss pattern as hydrogen atoms, which was attributed to strong site specificity of nonadiabatic effects.[19]

Naturally, the situation here is quite different from metals, where an EHP excitation is relatively easy due to the lack of a band gap, and as a result the perturbation to the nuclear motion can be approximated by the EF theory.[20-22] For the Ge surface, the inter-band transition across a large band gap of approximately 0.5 eV induced by the incidence of H/D requires a completely new theoretical framework. In the original publication,[17] the authors speculated that the non-adiabatic transition can be likened to a curve-crossing between two electronic states in a gaseous system, but the large number of electronic states in the VB and the CB need to be modeled if a first principles characterization is to be established. Currently, such a theory does not exist, although recent band structure calculations have provided some hints on the origin of the site specificity.[19] As depicted in Figure 1, there are three types of surface atoms on the reconstructed Ge(111), namely the adatoms, the rest atoms, and the saturated atoms, respectively. Their disparate coordination numbers make their interactions with the hydrogen atom distinct. It is thus intriguing to investigate the correlation between their local electronic structures and the non-adiabatic effects.

To this end, in this work, we analyze the change in the band structure of Ge(111)$c$(2×8) upon the adsorption of H on different surface sites and extract information about charge transfer and spin polarization. Furthermore, we follow explicitly the electronic dynamics during the direct scattering of H atoms off these sites by time-dependent non-adiabatic molecular dynamics (NAMD) simulations based on Kohn-Sham (KS) orbitals. While such simulations cannot model the experiment directly, clear and strong site-specific non-adiabatic effects are revealed, which support the curve-crossing hypothesis recently proposed.[17, 19] It is found that such curve-crossings are present when H atoms collide at the Ge rest atom site, most likely leading to inter-band transitions from the VB to the CB, while not at the Ge adatom site. The theoretical studies reported here thus advance our understanding of the newly discovered non-adiabatic effects in surface dynamics.

**II. Methods**

Spin-polarized plane-wave DFT calculations were performed with the Vienna Ab initio Simulation Package (VASP).[23] The electron-nuclei interaction was described by the projector augmented wave (PAW) method[24] and the KS wavefunction was expanded in plane waves with an energy cutoff of 400 eV. The reconstructed Ge(111) surface was represented by a slab in a $c(2\times8)$ supercell with the size of 8.0×32.0 Å plus a vacuum space of 16 Å in the $z$ direction to avoid artificial inter-slab interactions. Each slab contains eight atomic layers, with four additional Ge adatoms exposed on the top layer, as illustrated in Figure 1. The adatoms and top six layers were allowed to move, and the bottom layer was passivated by H atoms. A Gamma-centered $k$-point grid of (4 × 1 × 1) was used in geometric optimization and band structure calculations.

In our recent work,[17] the Perdew−Burke−Ernzerhof (PBE) density functional[25] within the generalized gradient approximation (GGA) was used to generate the ground state potential energy surface (PES), which well described the adiabatic energy transfer channel. However, we found that PBE improperly predicted a near-zero bandgap (~0.002 eV) for a clean surface. It is well-known that GGA-based density functionals often underestimate bandgaps of semiconductors owing to self-interaction errors.[26-27] While hybrid functionals provide generally more accurate bandgaps,[28-30] they are computationally much more expensive. Through tests over several commonly-used density functionals, we found that a meta-GGA made simple (MS2) functional[31] predicts a similar bandgap (0.32 eV) to that by the hybrid HSE06 functional[32] (0.35 eV) for the pristine Ge(111)$c$(2×8) surface, both close to the experimental value[18] (0.49±0.03 eV at 30 K), while the computational cost of the former is roughly one thousandth of the latter. In addition, the optimized lattice constant by MS2 is 5.688 Å, agreeing better with the experimental value (5.658 Å[33]) than the PBE value (5.783 Å). As a result, MS2 appears to provide a reasonably good description of the electronic structure of the system with a good balance between accuracy and efficiency, and is hence used in subsequent band structure and NAMD simulations.

A quantitative treatment of the non-adiabatic dynamics of surface scattering process is extremely challenging. Indeed, an accurate determination of a manifold of excited states of this periodic system is currently not feasible. Furthermore, the explicit evolution of nuclear and electronic degrees of freedom in a fully coupled way is even more formidable. In this work, we will focus on the response of the substrate electronic structure to the impact of the H atom at different surface sites, an approach which is aimed at uncovering the origin of the inter-band transition. The case of D scattering is not considered in this work, as the change of projectile mass is not expected to qualitatively alter the electron dynamics. A mixed quantum-classical strategy is used, in which the electronic degrees of freedom are treated quantum mechanically, while the nuclear motion classically. Specifically, we chose an approximate NAMD approach that solves the time-dependent Kohn-Sham (TDKS) equation at the one-electron orbital level based on the surface hopping algorithm[2] and the classical-path approximation (CPA).[34-36] This approach has been widely used in combination with ground-state ab initio molecular dynamics (AIMD) to study real-time photo-excited carrier dynamics in condensed phase materials.[37-41] Here, all NAMD calculations were performed with the Hefei-NAMD program.[39]

Specifically, the CPA assumes that the nuclear motion is governed by the ground state PES, unaffected by the electronic dynamics. In the TDKS/CPA approach, the electronic wavefunctions $\psi_e(\mathbf{r},t)$ is expanded in the basis of instantaneous adiabatic KS orbitals $\phi_j(\mathbf{r},\mathbf{R}(t))$ along a classical path of nuclei,

$$\psi_e(\mathbf{r},t) = \sum_j c_j(t)\phi_j(\mathbf{r},\mathbf{R}(t)). \qquad (1)$$

Here $\mathbf{r}$ and $\mathbf{R}$ represent the electronic and nuclear coordinates, respectively. Inserting Eq. (1) into the TD Schrödinger equation leads to a set of differential equations,

$$i\hbar \frac{\partial c_j(t)}{\partial t} = \sum_k c_k(t)\left[\varepsilon_k \delta_{jk} - i\hbar \mathbf{d}_{jk}(t)\right], \qquad (2)$$

where the squares of orbital coefficients, $|c_j(t)|^2$, correspond to the populations of the corresponding KS orbitals, $\varepsilon_k$ is the energy of the $k$th one-electron KS state, and $\mathbf{d}_{jk}$ is the time-dependent non-adiabatic couplings (NACs) between these KS states $j$ and $k$ defined by,

$$\mathbf{d}_{jk} = \left\langle \phi_j \left| \frac{\partial}{\partial t} \right| \phi_k \right\rangle = \sum_\mathbf{R} \frac{\left\langle \phi_j | \nabla_\mathbf{R} \mathcal{H} | \phi_k \right\rangle}{\varepsilon_k - \varepsilon_j} \cdot \dot{\mathbf{R}}. \qquad (3)$$

Apparently, time-dependent NACs are dependent on the energy difference $\varepsilon_k - \varepsilon_j$, the nuclear velocity $\dot{\mathbf{R}}$, and the derivative coupling vectors $\left\langle \phi_j | \nabla_\mathbf{R} \mathcal{H} | \phi_k \right\rangle$ that are calculated by a finite difference method.[42] The electronic dynamics along a given classical path of nuclei can be initiated from different electronic configurations, during which the single-electron hopping probabilities between these KS states can then be obtained according to Tully's fewest-switches algorithm,[2]

$$P_{j \to k}(t, \Delta t) = \frac{2\Re\left[c_j^* c_k \mathbf{d}_{jk}\right] \Delta t}{c_j^* c_j}, \tag{4}$$

where $\Re$ takes the real part of the quantity in the parenthesis. Since we are considering transient electronic excitations during a non-equilibrium collisional process, the Boltzmann factor to maintain the detailed balance was not imposed.[37] It is important to recognize that this ansatz does not properly describe the coupled electron-nuclear dynamics as there is no feedback to the nuclear motion from the electronic dynamics. As shown below, however, it is sufficient to qualitatively reveal the impact of the hyperthermal H on the substrate electronic structure and the potential mechanism for non-adiabatic transitions.

In practice, several representative MD trajectories on a previously constructed neural network PES[17] were selected to describe the H atom scattering off the Ge(111)$c$(2×8) surface at different surface sites at an incidence energy ($E_i$) of 0.99 eV, a representative $E_i$ used in the experiment[17]. The propagations of these trajectories were efficiently carried out using the high-dimensional NN PES. Along each trajectory, single point DFT calculations at the MS2 level were recomputed at selected snapshots every 0.5 fs to get the KS orbitals and associated time-dependent NACs. After that, NAMD trajectories were initiated with one electron filling in the valence band maximum (VBM) and then propagated along the selected nuclear trajectory with a step size of 1/1000 of that of AIMD, allowing electron excitation/de-excitation among 50 orbitals below and above the Fermi level. Due to the stochastic nature of the surface hopping algorithm, the final electronic state populations of different NAMD trajectories are not expected to be the same. To obtain the statistics, 20,000 NAMD trajectories were conducted for each site to yield the evolution of electronic populations in different orbitals, providing useful information on the electron dynamics.

## III. Results and Discussion

First, let us consider a static description of electronic structures of the Ge(111)$c$(2×8) surface with and without hydrogen adsorption. In a bare surface, as shown in Figure 1, the Ge adatoms (magenta) and the Ge rest atoms (yellow) are bound to three saturated atoms of the first layer (blue), both featuring a single dangling bond. The saturated atoms are tetrahedrally coordinated and have no unpaired valance electrons. Accordingly, the H atom preferably adsorbs on a Ge rest atom with a binding energy of 2.77 eV, followed by on a Ge adatom (2.25 eV), and on a Ge saturated atom (2.06 eV), at the MS2 level. Since the rest and adatoms pucker a little out of the surface and have a stronger attraction to the H atom, they cast a "shadow" over the saturated atoms, preventing the impinging H atom from directly colliding with latter to a large extent. Indeed, we find that merely 8% MD trajectories with our NN PES can be recognized as a single collision of the H atom at saturated atoms with $E_i$ = 0.99 eV, due to this "shadowing" effect, even though the proportion of saturated atoms is much higher than others on the surface. Therefore, we shall focus on the rest and adatoms subsequently.

Figure 2 compares the calculated band structures of the pristine Ge(111)$c$(2×8) surface, H adsorbed on a Ge adatom (H*@Ge(ad)), and on a Ge rest atom (H*@Ge(rest)) using the MS2 and HSE06 functionals, respectively. It is encouraging that the band structures obtained by the two density functionals are quite similar, except that the overall energy scale is slightly compressed in the MS2 cases. In both cases, the Ge(111)$c$(2×8) surface is predicted to have a finite bandgap or energy difference between VBM and the conduction band minimum (CBM), 0.32 eV for MS2 and 0.35 eV for HSE06, in reasonable agreement with the experimental value of 0.49 ± 0.03 eV.[18] This validates the accuracy of MS2 and all subsequent results are based on this functional. In addition, we find that the highest valance bands are mainly contributed by the rest atoms, while the lowest conductions bands are more related to the adatoms (and saturated atoms), more specifically, their $p_z$ orbitals.

We next discuss in more detail how the adsorption of an H atom influences the local electronic structure at different surface sites. When the hydrogen atom adsorbs, the original VB remains either at the H*@Ge(ad) or H*@Ge(rest). However, the lowest unoccupied bands are strongly influenced due to the formation of the new H-Ge bond, which can occur at both surface sites. When a hydrogen atom forms a chemical bond to a Ge adatom, the corresponding unoccupied $p_z$ orbitals of Ge atom hybridize with the 1s orbital of hydrogen atom and drop significantly in energy (out of the energy range in Figure 2), thereby leading to a lowering of the Fermi energy (Figures 2(b) and 2(e)) and making the VBM partially occupied. The relative downshift of Fermi energy is ~0.2 eV using MS2 and ~0.23 eV with HSE06. In contrast to this, H-Ge bond formation at the rest atom involves interaction with the fully occupied VB. Consequently, the hybridization of the H atom 1s orbital with the Ge (rest) orbitals, necessary to form a covalent bond, involves transfer of an electron to another originally unoccupied orbital of a dangling bond on the adatom, resulting in the presence of a new VBM.

This can be clearly seen in Figure 3, where the charge density difference between the pristine surface and the surface with an H atom bonding to an adatom and a rest atom is shown. Also shown in Fig. 3 are the corresponding spin density distributions. For hydrogen adsorption on the rest atom, it is evident that the H atom loses its electron, which transfers to an adatom next to the rest atom where the hydrogen atom attaches. This reverse charge transfer phenomenon was observed in earlier scanning tunneling microscope images.[43-44] More interestingly, since the H atom brings only one additional electron, the filled orbital of a dangling bond on the adatom (the new VBM) is present only in one spin manifold, as seen in Figures 2(c) and 2(f). It is further verified in Figure 3(d) that a strong spin density only appears on this adatom. This dramatic change sets the stage for the curve crossing discussed below. On the other hand, no such indirect charge transfer is observed when the H atom binds to the adatom, see Figure 3(a), suggesting that a crossing of electronic states is less likely (*vide infra*). It is shown that the H-Ge bond formed with the adatom enriches electrons on the H atom and borrows some electron density from nearby saturated Ge atoms. Additionally, there is no net spin density at this configuration (Figure 3(b)), confirming that the hydrogen adsorption on the adatom does not violate the spin-degeneracy, consistent with Figures 2(b) and 2(e). The comparison clearly underscores the drastic differences in the local electronic structures for H adsorption at different surface sites.

We next study the electronic dynamics of the hydrogen atom colliding at two surface sites of the Ge(111)*c*(2×8) surface by initially placing an electron in the VBM. Two exemplary scattering trajectories are illustrated in Figure 1. The corresponding time evolutions of the KS eigen-energies at the Γ point, the electronic populations of these eigenstates near the Fermi level, and the average electronic energy are shown in Figure 4. To simplify the problem, here the incident H atom is directed towards the specified site initially along the surface normal from 6 Å above the surface and the initial surface temperature is set to 0 K to avoid mixing of KS orbitals caused by the thermal motion of surface atoms. This model is by no means a faithful representation of the experiment, but it nevertheless provides insight into how the nuclear motion affects the electronic energy and transitions.

For the collision at the adatom site, as shown in Figures 4(a-c), the KS orbitals exhibited a synchronized lowering of their energies, as H approaches the surface reaching its distance of closest approach at about 25 fs. The system largely recovers when H departs after 40 fs. Importantly, overall, the order of the KS orbitals is almost kept the same and no near-degeneracies arise between VB and CB. The band gap remains throughout the scattering. As shown in Figure 4(b), the occupancy of VBM is kept at one, while all other orbitals are essentially unoccupied throughout, indicating that no electronic transition takes place. The average electronic energy, which is calculated by summing the energies of all electronic states weighted by their relative populations, shows two pronounced dips in Figure 4(c), reflecting the lowering of the KS energies (particularly that of VBM) as H forms a bond with the adatom. Overall, the incoming and outgoing average electronic energies are only slightly different due to the small energy change of the adiabatic KS orbitals, which is irrelevant to non-adiabatic transitions.

The situation is drastically different for H impinging on the rest atom. Note in this case that avoided crossings are present for one manifold of spin states, while absent in the other, as expected from the band structure of H@Ge(rest) in Figure 2 (where spin-orbit coupling is neglected). For the spin manifold with strong non-adiabatic characteristics (spin-down here), it is clear from Figure 4(d) that VBM first undergoes an avoided crossing with VBM-1 and VBM-2 at ~30 fs as the H projectile approaches the surface. Accordingly, as shown in Figure 4(e), the VBM occupancy starts to decrease, accompanied by a notable population in VBM-1 and a minor population in VBM-2. Immediately after, VBM steadily rises in energy by more than 0.3 eV at 33 fs, and more importantly, becomes near-degenerate with the CBM. This increase in energy is close to the band gap of the pristine Ge surface, leading to the absence of a band gap at the crossing point. The near-degeneracy reappears at 41 fs as H departs the surface and re-enters this curve-crossing area. In the meantime, there are some other avoided crossings between other KS orbitals, *e.g.*, CBM and CBM+1. As a result, we observe an increasing population in CBM and CBM+1 at 33 and 41 fs. Accordingly, the average electronic energy climbs dramatically as the H atom approaches the rest atom, as seen in Figure 4(f). In contrast, Figure 4(g) shows clearly in the other spin manifold (spin-up here), these KS orbital energy curves in the VB and CB do not cross at all. Consequently, electronic transitions are unlikely (Figure 4(h)) and the mean electronic energy varies only modestly (Figure 4(i)), analogously with the energy of VBM. We note however that the inclusion of the spin-orbit coupling may allow electronic transitions between different spin states.

We can further relate the site-specific electron dynamics of the H collision to the NACs along the NAMD trajectories. Figure 5 compares time-dependent and time-averaged NACs over all snapshots of NAMD trajectories for H scattering from the adatom and from the rest atom (for the spin-down manifold with strong couplings only). It is clear that for the H-atom collision at the adatom site, the KS orbitals above and below the Fermi level have negligibly small couplings. The largest NAC is found between VBM-3 and VBM-2 (see Figure 5(a)), which are deep in the VB and may cause some intra-band transitions, but should have little impact on the non-adiabatic inter-band transitions from VB to CB. Consequently, no electronic excitations from VB to CB are seen in this process. In contrast, for the H-atom collision at the rest atom, the NACs are on average much larger and more widely distributed among these orbitals near the Fermi level. Specifically, the NACs between VBM-1 and VBM, VBM and CBM, CBM and CBM+1, increase sharply when the KS orbitals undergo avoided crossings around 30 and 40 fs, indicating high probabilities of non-adiabatic transitions at these snapshots. This physical picture is fully consistent with that shown in Figure 4, corresponding to the changes of electronic populations in these orbitals. We note in passing that the NACs between CBM and CBM+1 are also large for the adatom site near 25 fs. While CBM is unoccupied at low temperatures, it can be thermally populated at elevated temperatures. Furthermore, one can also populate CBM by n-doping in a semi-conductor sample. In such cases, the transitions between CBM and CBM+1 could become relevant, for example, to recent surface temperature-dependent scattering experiments.[45]

It is worth mentioning that the transient near-degeneracies between VBM and CBM

observed here along the trajectories are analogous to avoided crossings between different electronic states in a gaseous system. The drastic changes in the KS energies are obviously caused by the strong perturbation of the electronic structure by the collision of the hyperthermal H atom. With sufficient collision energy, the impinging H pushes the system to reach a sufficiently high electronic energy, thus allowing the transition of a surface electron from the VB to the CB. The newly created high-energy EHP may dissipate its energy quickly to the Ge substrate and the rebounding H atom loses a significant portion of its kinetic energy, as observed in the experiment. Unfortunately, the latter scenario is not properly simulated within the current model, as the electronic dynamics does not alter the nuclear motion within the CPA. A more sophisticated theory is needed in the future.

## IV. Conclusions

Recent experimental exploration of gas-surface scattering has uncovered various non-adiabatic phenomena. Many such phenomena at metal surfaces can be framed within an electronic friction model in which thermal excitation of electron-hole pairs leads to a frictional force for the nuclear motion. However, such a friction model becomes inadequate for collisions from semi-conductor surfaces, which have significant band gaps. In recent experimental studies of H atom scattering from a reconstructed Ge(111)*c*(2×8) surface, significant energy losses were observed and attributed to a non-adiabatic transition of a surface electron from its valance band to conduction band. In this work, we present a mixed quantum-classical model to examine the collision induced non-adiabatic dynamics and its site specificity, in which the quantum electronic degrees of freedom are modeled by single-electron Kohn-Sham orbitals with classical nuclear motion within the classical path approximation.

Our non-adiabatic molecular dynamics simulations revealed significant collision-induced perturbation of the Kohn-Sham orbitals, leading to electronic transitions from the valance band to the conduction band. Specifically, the high kinetic energy of the impinging H atom enables the access of high-energy near-degeneracies between Kohn-Sham orbitals that are normally separated by the large band gap, resulting in the conversion of nuclear kinetic energy to electronic energy and ultimately the significant slowdown of the scattered H atom observed in the experiment. Such non-adiabatic transitions are possible at the rest atom site of the Ge surface and only in a particular spin manifold, but not at the adatom site, confirming our earlier speculations. This dramatic site-specificity in non-adiabatic dynamics can be traced back to the different local band structures perturbated by the approach of the energetic H atom at these sites. The theoretical insights shed valuable light on the non-adiabatic dynamics involving gas-surface interaction for large band-gap semiconductors. It also provides a useful construct for future development of more quantitative models for non-adiabatic processes on surfaces.

**Acknowledgements**: This work is mainly supported by the Strategic Priority Research Program of the Chinese Academy of Sciences (XDB0450101 to B.J. and J.Z.), National Natural Science Foundation of China (22325304 and 22221003 to B.J.), the US NSF (Grant Nos. CHE-1951328 and CHE-2306975 to H.G.) and the Alexander von Humboldt Foundation to H.G. O.B. and A.M.W. acknowledge support from the Deutsche Forschungsgemeinschaft (DFG) under Grant No. 217133147 (SFB1073, project A04) and from the DFG, the Ministerium für Wissenschaft und Kultur, Niedersachsen and the Volkswagenstiftung under Grant No. 191331650. A.M.W. thanks the Max Planck Society for the advancement of science. B.J. and H.G. also thank Prof. Wei Hu and Prof. Oleg Prezhdo, respectively, for some stimulating discussions. We thank the Supercomputing Center of USTC, Hefei Advanced Computing Center, Beijing PARATERA Tech CO., Ltd for providing high-performance computing service.(Grant Nos. CHE-1951328 and CHE-2306975 to H.G.) and the Alexander von Humboldt Foundation to H.G. O.B. and A.M.W. acknowledge support from the Deutsche Forschungsgemeinschaft (DFG) under Grant No. 217133147 (SFB1073, project A04) and from the DFG, the Ministerium für Wissenschaft und Kultur, Niedersachsen and the Volkswagenstiftung under Grant No. 191331650. A.M.W. thanks the Max Planck Society for the advancement of science. B.J. and H.G. also thank Prof. Wei Hu and Prof. Oleg Prezhdo, respectively, for some stimulating discussions. We thank the Supercomputing Center of USTC, Hefei Advanced Computing Center, Beijing PARATERA Tech CO., Ltd for providing high-performance computing service.

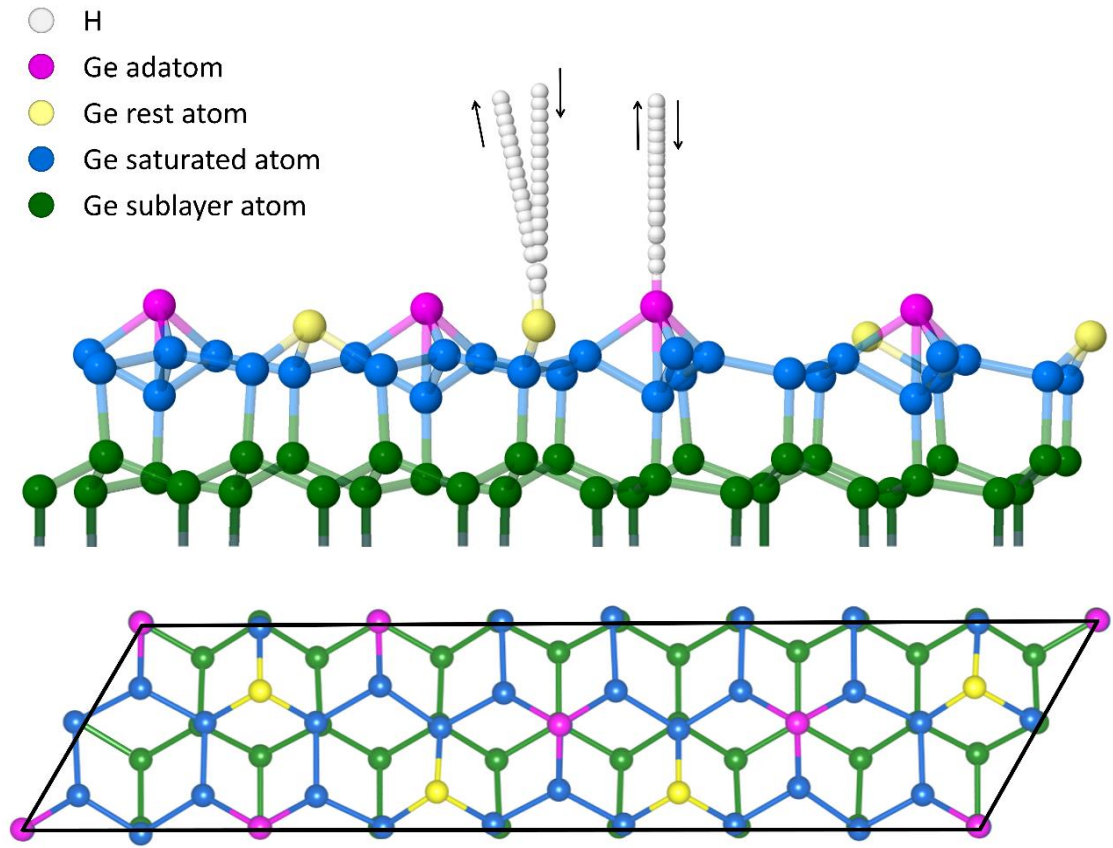

**Figure 1.** Schematics of two exemplary trajectories of H (white) collisions with a reconstructed Ge(111)*c*(2×8) surface at a Ge adatom (magenta) and a Ge rest atom (yellow) in a side view. A top view of the clean surface is also included for showing the site positions.

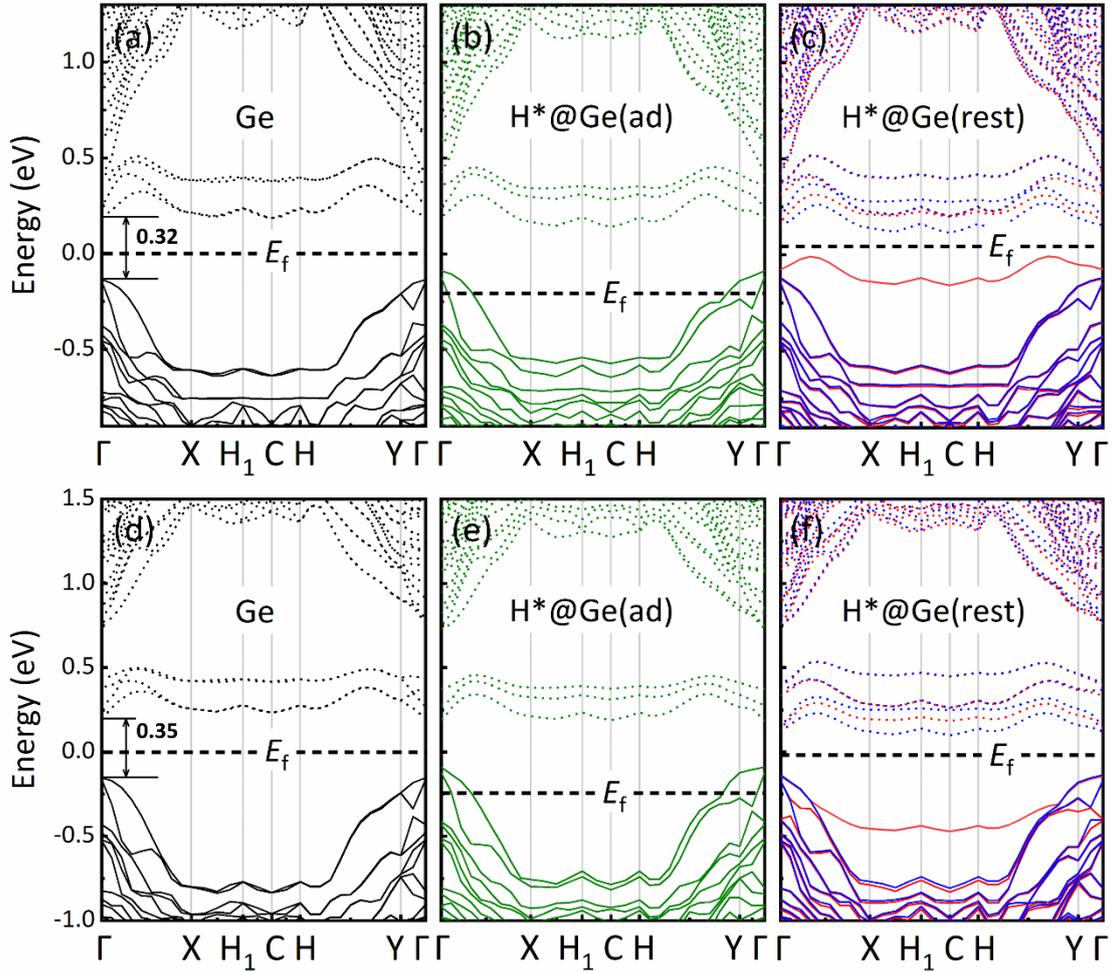

**Figure 2.** Comparison of the band structures of the Ge(111)$c$(2×8) surface with H(g) being 6 Å above (H(g)@Ge, left), H adsorbing on the Ge adatom (H*@Ge(ad), middle), and the Ge rest-atom (H*@Ge(rest), right), optimized by the MS2 (a-c) and HSE06 density functionals (d-f). Solid (dotted) lines represent occupied (unoccupied) energy levels. Fermi levels ($E_f$) are indicated by horizontal black dashed lines and the original band gaps of the Ge(111)$c$(2×8) surface are marked by vertical arrows (in eV). Spin-up and spin-down band structures are identical for the Ge and H*@Ge(ad) cases, but not for the H*@Ge(rest) case, which are distinguished by blue and red colors, respectively.

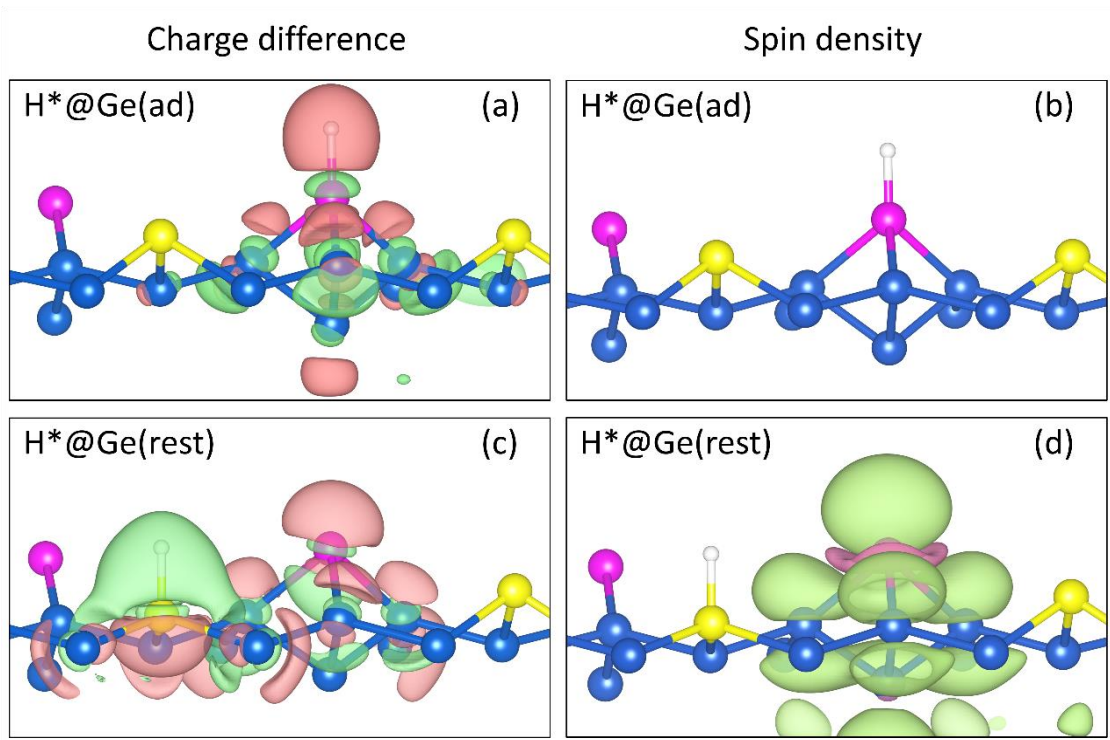

**Figure 3.** Differential charge density (a, c) and spin density (b, d) distributions of the H*@Ge(ad) and H*@Ge(rest) configurations. Negative values (gaining electrons or spin-down) are shown in pink and positive values (losing electrons or spin-up) in green. Iso-values are 0.007 eV/Å$^3$ for charge density and 0.0005 a.u. for spin density. Only the Ge atoms of the first layer and above are shown for clarity.

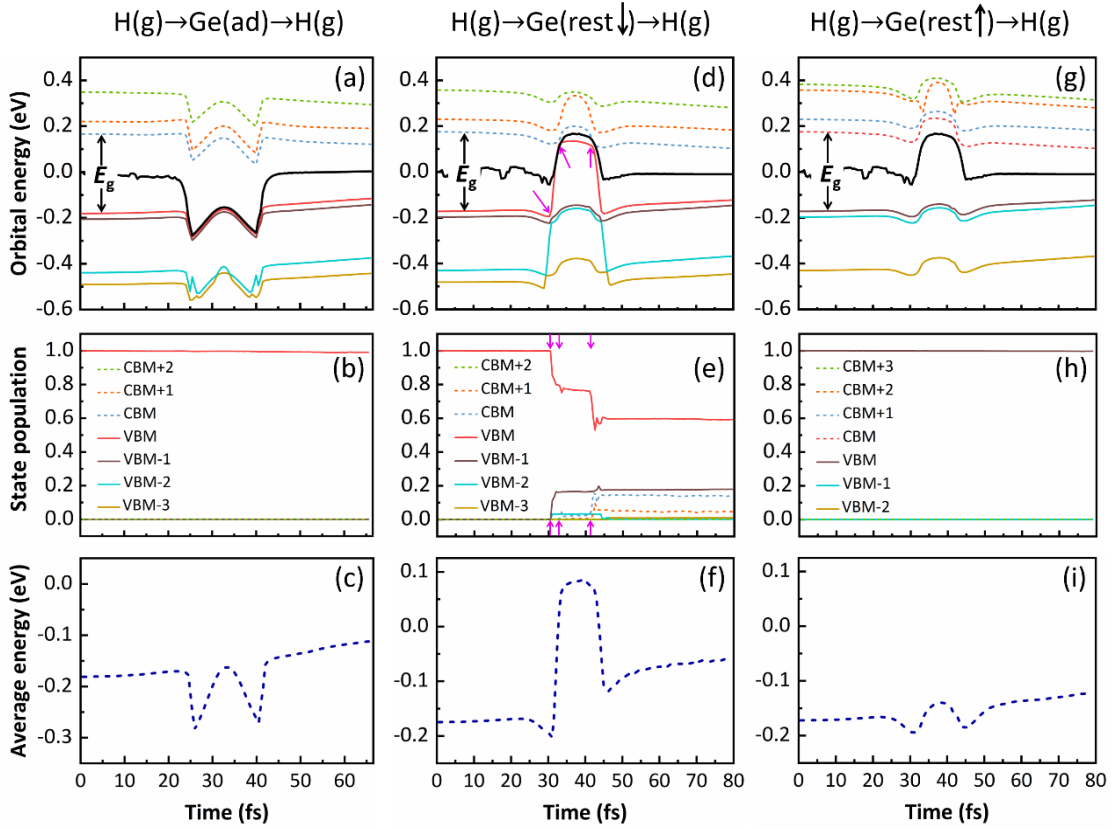

**Figure 4.** Time evolutions of the KS eigen-energies at the Γ point, the electronic populations of these energy levels near the Fermi level and the average electronic energy of the exemplary NAMD trajectories of H atom scattering from the adatom (a-c) and the rest atom (d-f for the spin-down and g-i for the spin-up eigenstates) of Ge(111)c(2×8) with an electron initially placed in the VBM and $E_i$=0.99 eV. Energy levels in the VB (CB) are given in solid (dotted) curves and labelled by their relative sequence to the VBM (CBM). In panels (a), (d), and (g), the initial band gaps ($E_g$) are marked by vertical arrows and the Fermi levels by the black solid lines. Pink arrows in panels (d) and (e) indicate the positions of avoided crossings where the electronic transitions occur.

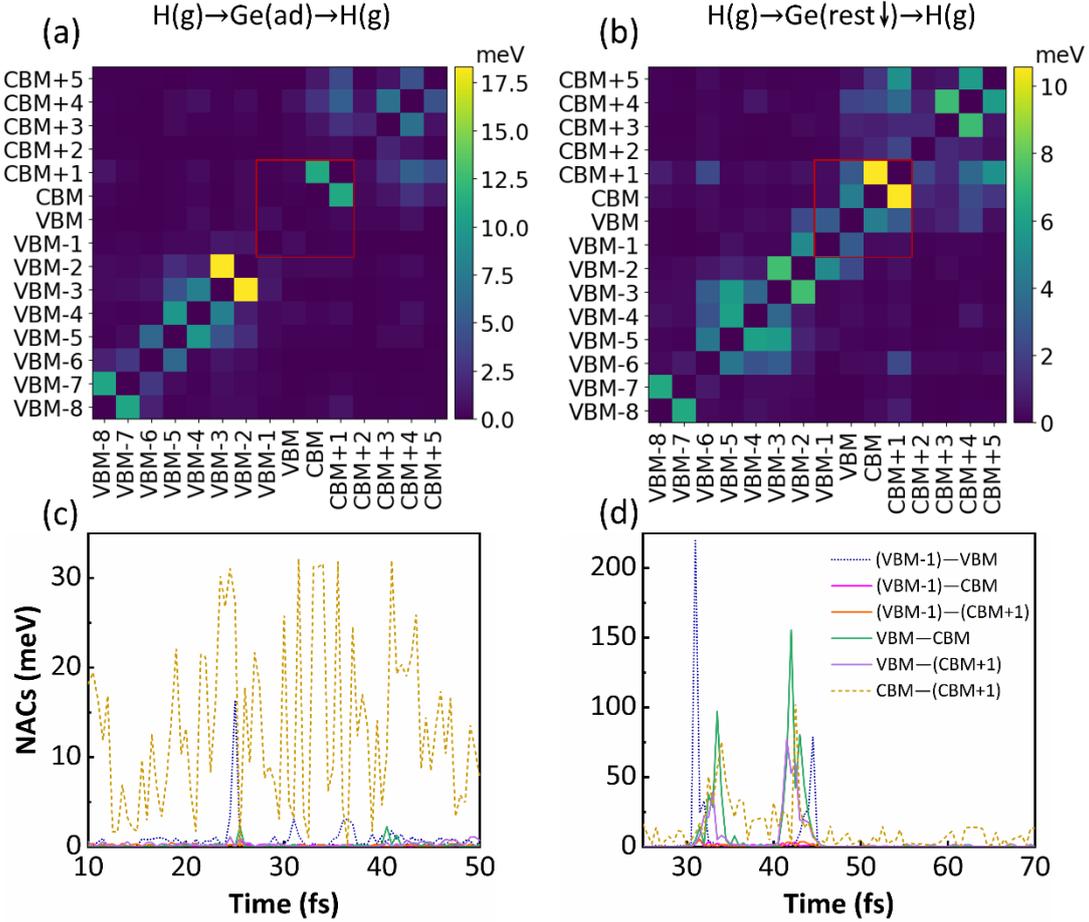

**Figure 5.** Time-averaged NAC matrices among different orbitals during the H atom scattering from the adatom (a) and rest atom (b) of Ge(111)*c*(2×8), where the orbitals are labelled by their sequence relative to VBM and CBM, respectively. NAC elements involving these orbitals near the Fermi level are highlighted in the red square and corresponding time-dependent values are presented in panels (c) and (d). Note for the rest atom that the NACs among spin-down eigenstates are shown only and all NACs are multiplied by $1000\hbar$ converting the unit to meV.